# A Comprehensive study of a New Multipath Energy Aware Routing Protocol for Mobile Ad-hoc Networks

S. Chettibi

Computer Sciences Dpt.,
University Mentouri of Constantine, Algéria.

*Abstract*—Maximizing network lifetime is a very challenging issue in routing protocol design for Mobile Ad-hoc NETworks (MANETs), since mobile nodes are powered by limited-capacity batteries. Furthermore, replacing or recharging batteries is often impossible in critical environments (e.g. battlefields, disaster areas, etc.)

The proposed MEA-DSR (Multipath Energy-Aware on Demand Source Routing) protocol uses a load distribution policy in order to maximize network lifetime. The simulation results have shown the efficiency of the proposed protocol in comparison to DSR routing protocol in many difficult scenarios.

## I. INTRODUCTION

A mobile ad-hoc network (MANET) is a collection of mobile devices communicating in a multi-hop fashion, without relying on any fixed infrastructure or centralized authority. MANETs are very attractive for applications where establishment of a communication infrastructure is impossible (e.g. in battlefields), or even when only a transient communication is needed (e.g. in conferences).

There are three current areas of research in energy efficient routing in ad-hoc networks [7]: 1) Power control, 2) Maximum lifetime routing and 3) Power save protocols.

In this paper, we present MEA-DSR a new multipath and energy aware on demand source routing protocol for MANETs and which belong to maximum lifetime routing class. The remainder of this paper is organized as follows: in section 2, we give motivations to our work and we describe in some detail MEA-DSR protocol. In section 3, we present and comment simulation results. Section 4 concludes the paper.

## II. MEA-DSR PROTOCOL

### A. Motivations

The intuition behind MEA-DSR protocol is that routes failure in MANETs is a norm rather than an exception. In conventional on demand routing protocols such as DSR [4] and AODV [2], a considerable amount of energy is lost due to frequent route discoveries. Furthermore, although frequent topology changes allow some kind of load distribution by forcing nodes to discover new routes but some nodes still used just because they have critical positions.

In MEA-DSR, node-disjoint multipath routing is exploited to: 1) minimise routes discoveries frequency since node disjoint routes are less likely to break simultaneously; 2) to balance energy consumption between mobile nodes. In addition, batteries-energy levels and paths length were also considered when making routing decisions.

### B. MEA-DSR protocol

MEA-DSR protocol is based on DSR. In fact, DSR is a multipath routing protocol. However, in DSR multiple routes are stored in a trivial manner with no constraint on number or quality.

The choice of the primary route in MEA-DSR is conditioned by two factors: 1) the residual energy of nodes belonging to the route; 2) the total transmission power required to transmit data on this route. If we consider that nodes transmit all with their maximal transmission power (adjusting transmission power feature is not supported by all network interface cards), the later factor is equivalent to that of hops number in the route. Concerning the choice of the second route, the disjunction ratio from the primary route comes in first order. If several routes present the same disjunction ratio, one will be chosen via the same criterion as for the primary route.

Instead of splitting traffic on several routes, only one route is used in MEA-DSR, during a communication session until its breakage. This permits to avoid problems of: route coupling [6], congestion of common nodes and out of order arrival of data packets to their destinations.

### C. Control packets and data structures used in MEA-DSR

Mobile nodes using MEA-DSR, exchange three types of control packets: route requests (RREQs), route replies (RREPs) and route errors (RERRs). The same format of RERR and RREP defined in [4] for DSR is used in MEA-DSR, whereas RREQ format has been slightly modified. MEA-DSR utilises three data structures: routes cache, route requests table and routes table. The same structure defined in [4] for routes cache is reutilised in MEA-DSR. Route requests table format has been enriched by additional fields; routes table is a new data structure specific to MEA-DSR.

*1. RREQ Packet*

A field called « min_bat_lev» has been added to RREQ packets. It takes as value the minimum of residual energies of nodes traversed by the RREQ packet.

*2. RREQs table*

The format defined in [4] has been augmented by the following fields: 1) «nb_hops» indicates the number of nodes traversed by the RREQ; 2) « last_node» maintains the identifier of the neighbor who transmitted the RREQ.

*3. Routes table*

This structure is utilised to store every candidate route in destination nodes, indexed by source node identifier. Every entry in the routes table contains the following fields:

- *Src*: maintains the identifier of source node who initiated the route discovery procedure.
- *Seq*: maintains the RREQ sequence number.
- *Route*: contains the nodes sequence traversed by RREQ packet.
- *Min_bat_lev*: keeps the minimal residual energy of nodes traversed by RREQ packet.
- *Arrival_time*: keeps the arrival time of RREQ packet at the destination node.

*C. Description of MEA-DSR protocol operation*

*1. Routes discovery*

If a source node needs a route toward a destination and no one is stored in its cache, it broadcasts a RREQ to all its neighbors. In MEA-DSR, only destination nodes can respond to a RREQ packet.

In order to avoid overlapped routes problem [5], intermediate nodes do not drop every duplicate RREQs and forward duplicate packets coming on a different link than the link from which the first RREQ is received, whose hop count is not larger than that of the first received RREQ. However, forwarding all duplicates satisfying this criterion generates a very high overhead. Thus, we have limited the number of copies to be forwarded to one.

When an intermediate node situated in the neighborhood of source node, receives the RREQ packet it includes its residual energy value in «min_bat_lev» field. Otherwise, any intermediate node compares its residual energy to the value of «min_bat_lev» field; if it is lower, it changes the value of «min_bat_lev» by its proper value. After the end of the above procedure, the intermediate node appends its identifier to the RREQ packet and rebroadcasts it to its neighbors. This process will be continued until that the RREQ packet arrives to destination.

*2. Routes selection*

After reception of the first RREQ packet, the destination node waits for a certain period of time "Wait_time" before starting route selection procedure. When this period of time expires, destination node selects as a primary route '$route_i$', satisfying the following condition:

$$\frac{\min\_bat\_lev_i}{route\_length_i} = \max_{j=1,n}\left(\frac{\min\_bat\_lev_j}{route\_length_j}\right) \quad (1)$$

Where, n is the number of candidate routes stored in the routes table.

After primary route selection, destination node sends immediately a route reply to the source node. The alternative route must be maximally node disjoint than the primary route. If there exist several routes with a same disjunction ratio, one that satisfies equation (1) will be chosen.

*3. Routes maintenance*

In MEA-DSR, if an intermediate node detects a link failure, it transmits a RERR message to the upstream direction of the route. Every node receiving the RERR message, removes every entry in its route cache that uses the broken link, and forwards RERR message to the next node toward the source node. If the source node has no valid route in its cache, then it reinitiates a new route discovery cycle.

## III. SIMULATION

**Table 1.** Studied simulation scenarios

| | Mobility | Density | Traffic |
|---|---|---|---|
| **Pause time** | Varied from 0 to 600s | 100 s | 100 s |
| **Maximum speed randomly chosen from the interval** | Three cases were considered: a) Low speed: [0,5 m/s, 1 m/s]. b) Moderate speed: [5 m/s, 10 m/s]. c) High speed: [20 m/s, 25 m/s]. | [5m/s,10m/s] | [5m/s, 10m/s] |
| **Nodes number** | 50 | Varied from 50 to 100 | 50 |
| **Data send rate and number of data sessions** | 4 pkt/s with 10 data sessions | 4 pkt/s with 10 data sessions | a) Data send rate varied from 2pkt/s to 12pkt/s with 10 data sessions b) Data send rate fixed at 4pkt/s and number of data sessions varied from 10 to 40. |

*A. Simulation environment*

NS-2 simulator [8] was used for MEA-DSR and DSR performances comparison. The studied network is deployed on square area of 1000mx1000m. Each node has a transmission range of 250 m. The MAC protocol was based on IEEE 802.11 with 2 Megabits per second raw capacity. For radio propagation model, a two-ray ground reflection model was used. In all simulations, we have utilized the RWP mobility model [1]. The duration of every simulation was 600 seconds. Communication between nodes was modelled by CBR traffic over UDP. Source nodes generate packets of 512 bytes. Generated connections start at a time randomly chosen from the interval [0s, 120s] and still active until the end of simulation.

Since we did not address the problem of consumed energy in idle state we have only considered energy consumed in transmission and reception modes. As values, we have utilised those obtained trough experiments in a previous work [3] (*1.4 W for transmission mode and 1 W for reception mode*).

To study mobility, density and traffic impact on MEA-DSR and DSR performances we have considered several scenarios as illustrated in Table 1.

*B. Performance metrics*

Studied performance metrics are the following:
- Normalized routing overhead (NRO)- the ratio of the number of routing protocol control packets transmitted to the number of data packets well received by destination nodes.

- Packet delivery fraction (PDF)- the ratio of data packets well received to those generated by source nodes.
- Consumed energy per packet (CEP)- the ratio of global consumed energy to the number of data packets well received.
- Standard deviation of consumed energy per node (SDCEN) - square root of the average of the squares of the difference between the energy consumed at each node and the average energy consumed per node.
- Minimal residual energy ratio (MRER) - the minimal of nodes residual energy to the initial energy ratio.

### C. Simulation results

#### 1. Node mobility impact on MEA-DSR and DSR performances

In all what follow, we mean by high mobility (low mobility): a high speed (low speed) and/or a frequent mobility (less frequent mobility) which is function of pause time duration.

##### 1.1. Normalized routing overhead

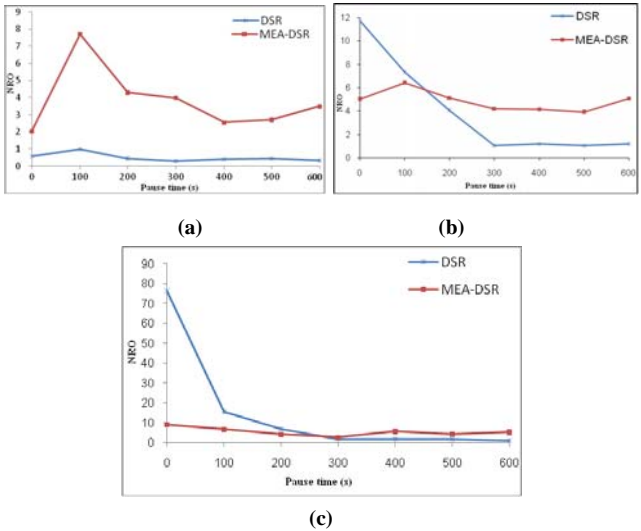

**Figure 1.** NRO vs pause time in a) case of low speed b) moderate speed c) high speed.

Under high mobility scenarios, DSR generates more routing overhead than MEA-DSR, because DSR reinitiates route discoveries repeatedly and consequently generates more RREQs.

In lower mobility scenarios, routes become more stable. Thus, the need to reinitiate new route discoveries is diminished for both protocols. However, MEA-DSR stills generate high overhead. This is because MEA-DSR permits intermediate nodes to propagate RREQs duplicates, whereas in DSR intermediate nodes drop every duplicate RREQ. Furthermore, RREQ packets in MEA-DSR always propagate until their final destination, whereas in DSR intermediate nodes can directly reply from their caches.

##### 1.2. Packet delivery fraction

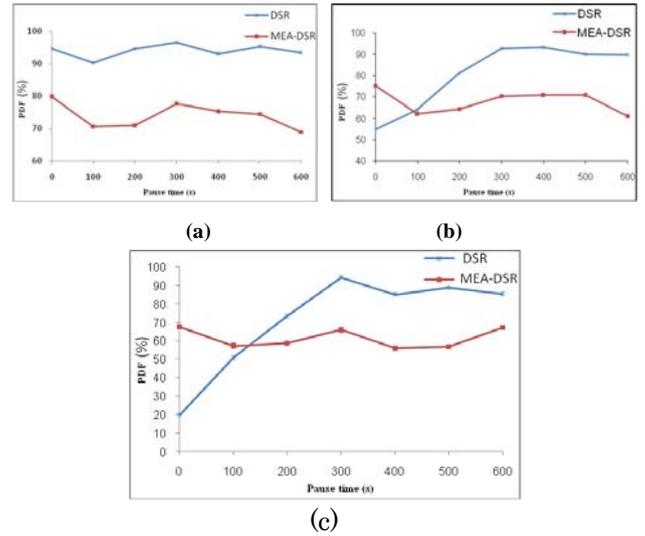

**Figure 2.** PDF vs Pause time in a) case of low speed b) moderate speed c) high speed.

Under high mobility scenarios, MEA-DSR presents a higher PDF than DSR. The reason is that intermediate nodes in DSR, are authorized to answer from their caches. However, in such mobility conditions stored routes are more likely to be stale. Thus, data packets forwarded on those routes will be dropped, as soon as they reach broken links, since also salvaging mechanism becomes less efficient. In addition, data packets are more likely to expire because of the additional latency introduced by frequent retransmissions and repetitive salvaging attempts. For lower mobility scenarios, PDF of DSR increases to surpass MEA-DSR one, because routes tend to be more robust and both responding from cache and salvaging mechanisms become more efficient. In MEA-DSR, intermediate nodes are not authorized to use their caches to salvage data packets. Thus, data packets probability to be dropped is greater than it is in DSR.

##### 1.3. Consumed energy per packet

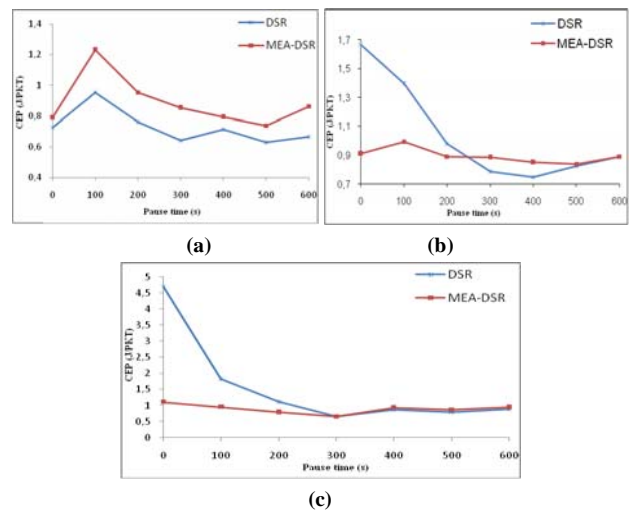

**Figure 3.** CEP vs Pause time in a) case of low speed b) moderate speed c) high speed.

CEP gives an idea about the global energy consumption in the network; it is clear that is proportional to both generated routing overhead and utilized routes length. For high mobility scenarios, DSR generates more overhead than MEA-DSR. Thus, it consumes more energy. For lower mobility scenarios, although DSR generates less overhead but it does not present an important improvement in energy consumption because it tends to use longer routes (total transmission power of a packet stills high).

*1.4. Standard deviation of consumed energy per node*

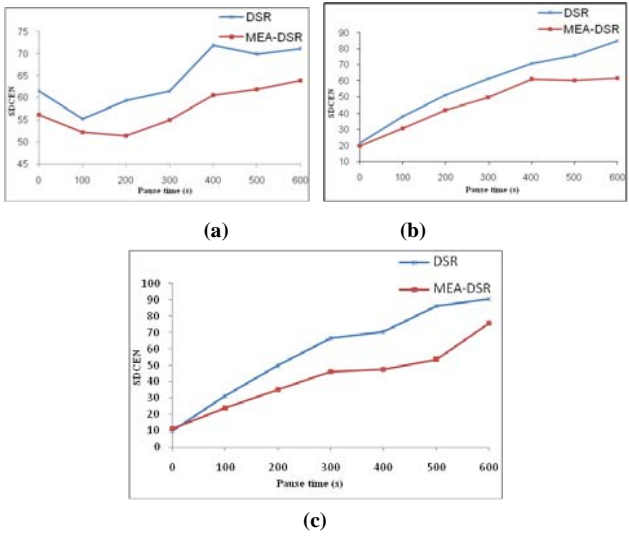

**Figure 4.** SDCEN vs Pause time in a) case of low speed b) moderate speed c) high speed.

Under all mobility scenarios, SDCEN in MEA-DSR is always lower than that of DSR, which confirms the efficiency of load distribution policy adopted in MEA-DSR.

For both protocols, network stability provokes SDCEN increase. This was expected, since routes still in use in a communication session while they are valid.

*1.5. Minimal residual energy ratio*

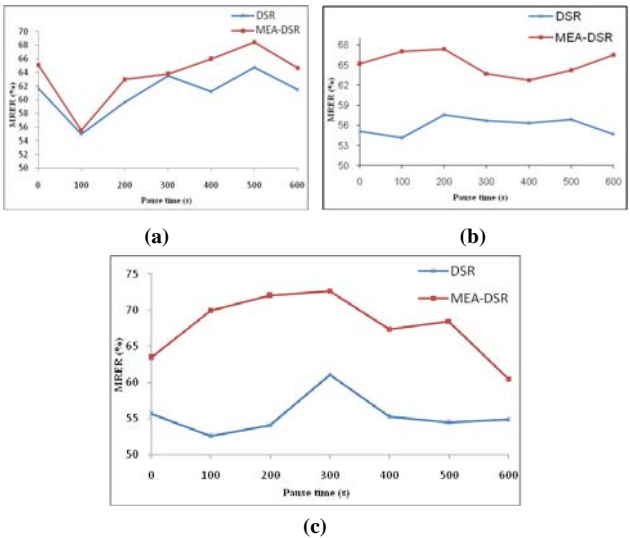

**Figure 5.** MRER vs Pause time in a) case of low speed b) moderate speed c) high speed.

Since MEA-DSR was fairer than DSR, under all mobility scenarios, its MRER was the highest. Under high mobility scenarios, the gain in residual energy was more important thank to its lower global energy consumption.

*2. Load traffic impact on MEA-DSR and DSR performances*

*2.1 Normalized Routing Overhead*

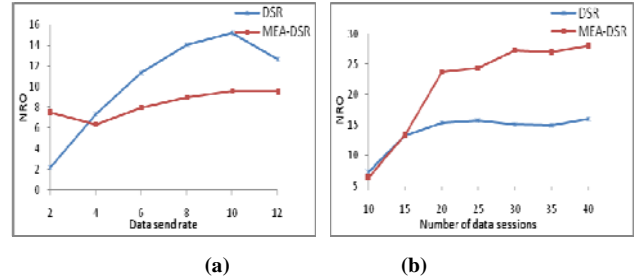

**Figure 6.** NRO vs a) data send rate b) number of data sessions

If we increase data send rate, more routes breaks will occur involving more routes reconstructions. If we increase number of data sessions more protocol operations will be initiated. In either case, this will increase the generated routing overhead.

When augmenting data send rate, NRO of MEA-DSR was less important than DSR one. This is because DSR needs to reinitiate routes discoveries more frequently than MEA-DSR per data session. When augmenting number of data sessions, NRO of MEA-DSR was more important than DSR one. This is normal since the number of exchanged RREQs per route discovery cycle in MEA-DSR is greater than in DSR.

*2.2. Packet delivery fraction*

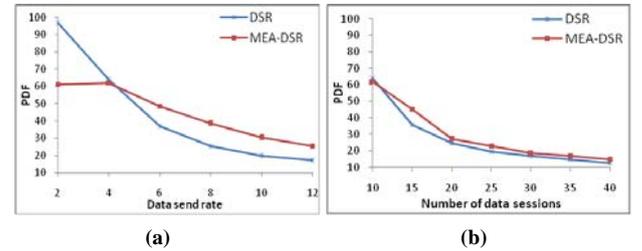

**Figure 7.** PDF vs a) data send rate b) number of data sessions

The augmentation of traffic load increases the risk of congestion and interferences which provokes more packets loss. This explains why the PDF of both protocols have decreased while increasing number of data sessions and data send rate. When increasing data send rate, MEA-DSR outperformed DSR. This is because MEA-DSR suffers less from queue congestion since it generates less overhead per data session than DSR. Although increasing number of data sessions has provoked an augmentation of the generated overhead by MEA-DSR, but DSR stills drop more data packets than MEA-DSR due to queue congestion.. This behavior can be explained by the fact that DSR spends more time before liberating an interface-queue entry corresponding to a non-acknowledged packet by making several salvaging attempts (under high mobility conditions cached routes are more likely to be stale). During this time, new arriving data packets will be dropped. Particularly, this is problematic when packets transmitters do not maintain alternate routes or do not maintain valid ones.

*2.3. Consumed energy per packet*

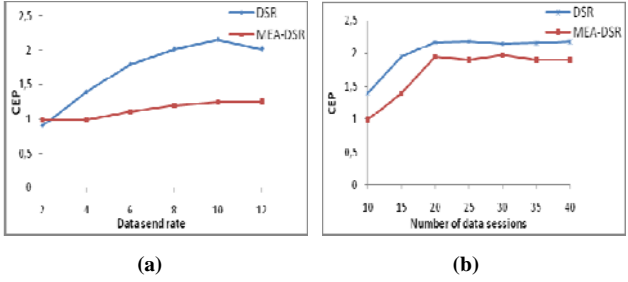

(a) (b)

**Figure 8.** CEP vs a) data send rate b)number of data sessions

When augmenting data send rate, CEP of MEA-DSR was inferior from DSR one. This is because MEA-DSR generates less routing overhead than DSR. Although when increasing number of data sessions, the routing overhead generated by DSR was less important than that generated by MEA-DSR, but this did not yield to DSR outperformance. This can be explained by the fact that DSR consumes more energy than MEA-DSR due to its ineffective salvaging attempts.

*2.4. Standard deviation of consumed energy per node*

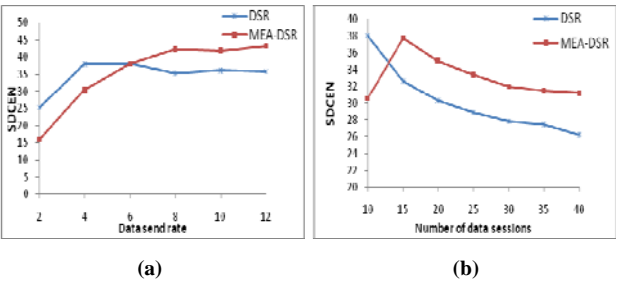

(a) (b)

**Figure 9.** SDCEN vs a) data send rate b)number of data sessions

The efficiency of load distribution strategy implemented in MEA-DSR was clear in cases of 2 and 4 pkts/s send rates. In all other cases (data send rate from 8 to 12 pkt/s and number of data sessions from 15 to 40), SDCEN of MEA-DSR was the worst. This is can be explained by the reduction in number of RREQS received at destination nodes due to interferences and congestion provoked by traffic load augmentation. When increasing data send rate from low to moderate values (2 to 6 pkt/s), SDCEN of both protocols have increased since more traffic is injected between same source-destination pairs. Hence, same nodes are utilized more frequently. When increasing the number of data sessions, both protocols have marked an enhancement in their SDCEN. This is because more nodes initiate communications, and thus will have very similar energy-consumption profiles.

*2. 5. Minimal residual energy ratio*

When increasing data send rate, MRER of MEA-DSR was more important than DSR since the global energy consumption in MEA-DSR was lower than DSR one. This was also true in cases of 10 and 15 data sessions. From 20 to 40 data sessions, the MRER of DSR was somewhat enhanced thanks to its nodes usage fairness mitigation.

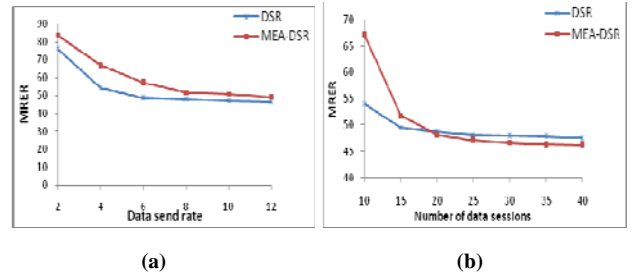

(a) (b)

**Figure 10.** MRER vs a) data send rate b)number of data sessions

## 3. Node density impact on MEA-DSR and DSR performance

*3.1. Normalized routing overhead*

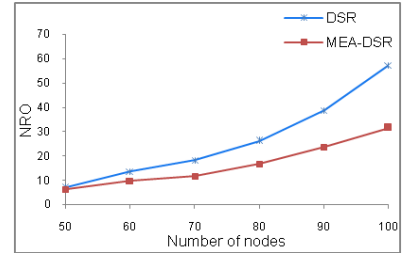

**Figure 11.** NRO vs number of nodes

As expected, the NRO of both protocols increased with node density growth. It is clear that under all density conditions MEA-DSR NRO was inferior from DSR one. This is because MEA-DSR reinitiates routes discoveries less frequently than DSR thanks to the use of maximally node-disjoint routes, and thus generates less control packets.

*3.2. Packet delivery fraction*

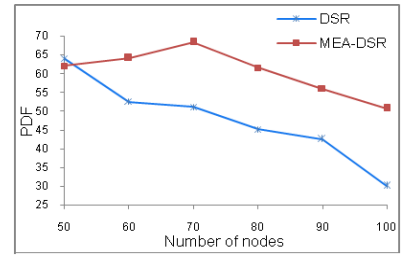

**Figure 12.** PDF vs. number of nodes

Under all density conditions (expect for the case of 50 nodes where PDF of both protocols was approximately the same) the PDF of MEA-DSR was higher than DSR one. This can be explained by the fact that DSR suffers more than MEA-DSR from queue congestion problem since it generates more overhead. Furthermore, routing overhead for both protocols increases with node density which had aggravated queue congestion problem. This explains the reduction of both protocols PDF according to node density augmentation.

By examining trace files of simulations, we found that MEA-DSR tends to use longer routes when node density increases which lead to an augmentation of routes failure probability. This is another raison for MEA-DSR PDF decrease.

### 3.3. Consumed energy per packet

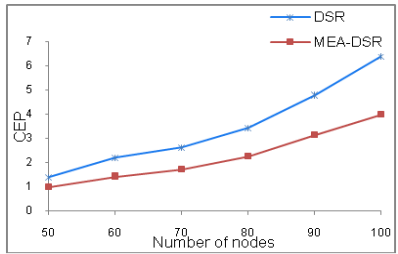

**Figure 13.** CEP vs number of nodes

Increasing node density leads to an augmentation of collisions risk (consequently to more retransmission attempts) and to a growth in number of exchanged control packets. All those factors cause more energy dissipation for both protocols. MEA-DSR CEP was lower than DSR CEP under all density conditions because MEA-DSR always generates less overhead than DSR. Thus, its global energy consumption remains lower than DSR one.

### 3.4. Standard deviation of consumed energy per node (SDCEN)

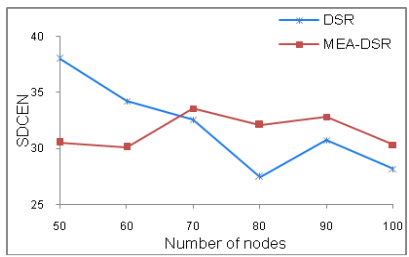

**Figure 14.** SDCEN vs number of nodes

The efficiency of load distribution policy implemented in MEA-DSR was clear in cases of 50 and 60 nodes, where the SDCEN of MEA-DSR was inferior from DSR one. Changing node density had not a significant impact on SDCEN of MEA-DSR; it stilled varying from 30 to 33. The load distribution in DSR was enhanced with node density increase. This is due to the variety of routes replies received from intermediate nodes (*increasing node density leads to an augmentation of neighbors number per node and thus to an increase in RREPs variety*).

### 3.5. Minimal residual energy ratio

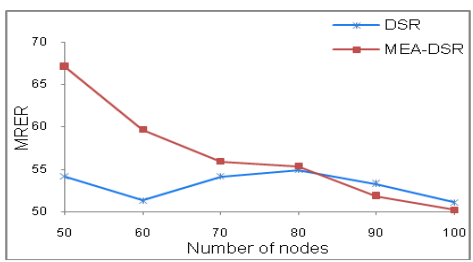

**Figure 15.** MRER vs number of nodes

Since the SDCEN and thus fairness of MEA-DSR was approximately the same under all density conditions MEA-DSR MRER diminution according to node density augmentation is justified by the increase of generated routing overhead.

Although the generated routing overhead had also increased in DSR, but this did not lead to a reduction of its MRER. In fact, under all density conditions DSR MRER was nearly the same and it varied between 51% to 54%. This result can be explained by the enhancement of nodes usage fairness of DSR. Nevertheless, DSR MRER was low in approximately all cases in comparison to MEA-DSR since DSR generates typically more routing overhead than MEA-DSR.

## IV. CONCLUSIONS

Maximally node-disjoint routes are exploited in MEA-DSR to 1) achieve a global energy gain by minimizing routes discoveries frequency; and to 2) balance energy consumption between mobile nodes. The choice of the primary route in MEA-DSR is dictated by minimal residual node energy to route length ratio, whereas disjunction ratio from primary route comes in first order in alternate route choice. Simulation results have shown energy efficiency of the proposed protocol under difficult scenarios characterized by high mobility, high density and important traffic load.

It is worth to notice that the efficiency of load sharing strategy implemented in MEA-DSR is directly influenced by number of RREQs arriving to destination nodes. This later is determined by congestion and interferences level present in the network.

One drawback of our protocol is its relatively low PDF. In a future work, we plan to investigate the impact of equipping intermediate nodes with alternate maximally node disjoint routes on both network energy consumption and throughput.